\documentclass[a4paper,11pt]{article}

\usepackage{jheppub}

\usepackage{amsmath,color}  

\def\prg#1{\medskip\noindent{\bf #1}}  
        
              \def\pd{\partial}

\def\mb{{\rm MB}}                \def\ct{{\rm ct}}
    \def\ads3{{\rm AdS$_3$}}
\def\Leff{\hbox{$\mit\L_{\hspace{.6pt}\rm eff}\,$}}

        \def\S{\Sigma}        \def\L{{\mit\Lambda}}
\def\D{\Delta}        \def\Th{\Theta}
\def\a{\alpha}        \def\b{\beta}         \def\g{\gamma}
\def\d{\delta}        \def\m{\mu}           \def\n{\nu}
\def\th{\theta}       \def\k{\kappa}        \def\l{\lambda}
    \def\ve{\varepsilon}  
\def\r{\rho}                 \def\om{\omega}
\def\s{\sigma}        \def\t{\tau}          
\def\nab{\nabla}

   \def\bg{{\bar g}}     
\def\cL{{\cal L}}     \def\cM{{\cal M }}    
\def\cH{{\cal H}}        
        
\def\cO{{\cal O}}      
     \def\cA{{\cal A}}   \def\hcA{\hat{\cal A}}

\def\bom{{\bar\omega}}    \def\tom{{\tilde\omega}}
\def\bare{{\bar e}}       \def\te{{\tilde e}}
\def\hnab{{\hat\nabla}}
\def\hT{{\hat T}}     \def\hR{{\hat R}{}}
\def\he{{\hat e}}     \def\hxi{\hat{\xi}}     \def\hth{\hat{\theta}}
\def\hg{\hat{g}}      \def\hom{\hat{\omega}}

\def\ren{{\rm ren}}

\def\nn{\nonumber}
\def\be{\begin{equation}}             \def\ee{\end{equation}}
\def\ba#1{\begin{array}{#1}}          \def\ea{\end{array}}
\def\bea{\begin{eqnarray} }           \def\eea{\end{eqnarray} }
\def\beann{\begin{eqnarray*} }        \def\eeann{\end{eqnarray*} }
\def\beal{\begin{eqalign}}            \def\eeal{\end{eqalign}}
\def\lab#1{\label{eq:#1}}             \def\eq#1{(\ref{eq:#1})}
\def\bsubeq{\begin{subequations}}     \def\esubeq{\end{subequations}}
\def\bitem{\begin{itemize}}           \def\eitem{\end{itemize}}

\title{\boldmath Holography in 3D AdS gravity with torsion}

\author[a]{Milutin Blagojevi\'c,}
\author[a]{Branislav Cvetkovi\'c,}
\author[b]{Olivera Miskovic}
\author[c]{and Rodrigo Olea.}

\affiliation[a]{Institute of Physics, University of Belgrade, \\
P. O. Box 57, 11001 Belgrade, Serbia}
\affiliation[b]{Instituto de F\'\i sica, Pontificia Universidad
            Cat\'olica de Valpara\'\i so, \\ Casilla 4059, Valpara\'{\i}so, Chile }
\affiliation[c]{Universidad Andres Bello, Departamento de Ciencias
           F\'\i sicas, \\ Rep\'ublica 220, Santiago, Chile }

\emailAdd{mb@ipb.ac.rs}
\emailAdd{cbranislav@ipb.ac.rs}
\emailAdd{olivera.miskovic@ucv.cl}
\emailAdd{rodrigo.olea@unab.cl}

\abstract{Basic aspects of the AdS/CFT correspondence are studied in the
framework of 3-dimensional gravity with torsion. After choosing a
consistent holographic ansatz, we formulate an improved approach to the
Noether--Ward identities for the boundary theory. The method is applied
first to the topological Mielke--Baekler model, and then to the more
interesting (parity-preserving) 3-dimensional gravity with propagating
torsion. In both cases, we find the finite holographic energy-momentum
and spin currents and obtain the associated (anomalous) Noether--Ward
identities.}


\begin{document}
\maketitle
\flushbottom

\section{Introduction}
\setcounter{equation}{0}

According to the idea of AdS/CFT correspondence \cite{x1}, to
any asymptotically anti-de Sitter (AdS) gravitational theory on
a $(d+1)$-dimensional spacetime $M$, there corresponds a
$d$-dimensional conformal field theory (CFT) on the boundary
$\pd M$. This duality is of the weak/strong coupling type: the
weak coupling regime of the gravitational theory is related to
the strong coupling regime of the boundary CFT, and vice versa.

Following a widely spread belief that general relativity (GR) is
the most reliable approach for studying the gravitational
phenomena, the analysis of the AdS/CFT correspondence has been
carried out mostly in the realm of Riemannian geometry, leading
to a number of highly interesting results \cite{x2,x3}. However,
one should note that, for nearly five decades, there exists a
modern gauge-theoretic conception of gravity, characterized by a
Riemann-Cartan geometry of spacetime. In this approach, known as
Poincar\'e gauge theory (PGT) \cite{x4,x5,x6}, both the torsion
and the curvature carry the gravitational dynamics. In spite of
its well-founded dynamical structure, the use of this framework
for studying the AdS/CFT correspondence is still in a rather
rudimentary phase. In this regard, we wish to mention the work
of Ba\~nados et al. \cite{x7}, who studied the holographic
currents in the 5-dimensional (5D) Chern--Simons gravity with
torsion, and the paper of Klemm and Tagliabue \cite{x8},
investigating the holographic structure of the Mielke--Baekler
(MB) model of 3D gravity with torsion \cite{x9}. In 4D, Petkou
\cite{x10} examined holographic aspects of Einstein--Cartan
theory amended by topological torsional invariants.

In order to properly understand dynamical features of gravity
with torsion, one is naturally led to consider technically
simplified models with the same conceptual features. An
important and useful model of this type is the MB model of
topological 3D gravity with torsion \cite{x9}, introduced in the
early 1990s. Further investigations along these lines led to a
number of remarkable results; for more details, see
\cite{x11,x12} and references therein. Of particular interest
for our present work is the existence of a holographic
structure, as discussed in \cite{x8}. However, in the MB model
(like in GR with a cosmological constant) there are no
propagating degrees of freedom. In order to overcome this
unrealistic feature of the gravitational dynamics, a systematic
study of 3D gravity with \emph{propagating} torsion has been
recently initiated in \cite{x12}, see also \cite{x13}. The
present work is aimed at investigating holographic aspects of 3D
gravity with (propagating) torsion, in order to reexamine the
compatibility of the concept of torsion with the basic aspects
of the AdS/CFT correspondence, and moreover, to understand the
dynamical role of the new CFT sources associated with torsion.

The paper is organized as follows. In section 2, we discuss
general holographic features of 3D gravity with torsion,
\emph{with or without} the propagating torsion modes. After
choosing a suitable ansatz for the gravitational variables,  we
derive the related consistency conditions, which tell us that
the maximal boundary symmetry consists of the local Poincar\'e
transformations and dilatations. In section 3, we propose an
improved treatment of the corresponding Noether--Ward identities
for the boundary theory. In section 4, we use this approach to
reexamine the holographic structure of the \emph{topological} 3D
gravity with torsion; our results confirm the analysis of Klemm
and Tagliabue \cite{x8}, based on a different technique. Then,
in section 5, we turn to the main subject of the present
work---the study of holography in 3D gravity with
\emph{propagating} torsion. We find that the maximal boundary
symmetry is reduced by the existence of the conformal anomaly.
The improved formalism ensures that these results do not depend
on the value of torsion on the boundary.

Our conventions are given by the following rules. In 3D
spacetime $M$, the Latin indices $(i,j,k,\dots)$ refer to the
local Lorentz frame, the Greek indices $(\m,\n,\r,\dots)$ refer
to the coordinate frame, the metric components in the local
Lorentz frame are $\eta_{ij}=(+1,-1,-1)$, totally antisymmetric
tensor $\ve^{ijk}$ is normalized by $\ve^{012}=+1$, and
symmetric and antisymmetric pieces of a tensor $X_{ij}$ are
$X_{(ij)}=\frac{1}{2}(X_{ij}+X_{ji})$ and
$X_{[ij]}=\frac{1}{2}(X_{ij}-X_{ji})$, respectively. Next, the
$(1+2)$ decomposition of spacetime is described in terms of the
suitable coordinates $x^\m=(\r,x^\a)$, where $\r$ is a radial
coordinate and $x^\a$ are local coordinates on the boundary $\pd
M$; in the local Lorentz frame, this decomposition is expressed
by $i=(1,a)$. Then, on 2D boundary $\pd M$ (which is orthogonal
to the radial direction), we have $\eta_{ab}=(+1,-1)$ and
$\ve^{ab}:=\ve^{a1b}$, with $\ve^{02}=+1$. Finally, we use the
Stokes theorem in the form $\int\pd_\l V^\l d^3 x=\int V^\r d^2
x$, where $V^\l=(V^\r,V^\a)$ is a vector density,  and $d^3 x$
and $d^2 x$ are coherently oriented volume forms on $M$ and $\pd
M$, respectively.

\section{Holographic ansatz}
\setcounter{equation}{0}

In this section, we introduce a general setting for 3D gravity with
torsion and discuss a suitable holographic ansatz for the basic
dynamical variables.

Three-dimensional gravity with torsion can be naturally described in
the framework of PGT \cite{x11,x12}, where basic gravitational
variables are the triad field $\he^i$ and the Lorentz connection
$\hom^{ij}=-\hom^{ji}$ (1-forms), the corresponding field strengths are
$\hT^i=d\he^i+\hom^i{_j}\wedge\he^j$ and
$\hR^{ij}=d\hom^{ij}+\hom^i{_k}\wedge\hom^{kj}$ (2-forms), and the
covariant derivative $\hnab=d+\frac{1}{2}\hom^{ij}\S_{ij}$ (1-form)
acts on local Lorentz spinors/tensors in accordance with their
spinorial structure, encoded in the form of the spin matrix $\S_{ij}$.
The antisymmetry of $\hom^{ij}$ ensures that the underlying geometric
structure of spacetime is given by the Riemann--Cartan (RC) geometry,
in which $\he^i$ is an orthonormal frame,
$\hg=\eta_{ij}\he^i\otimes\he^j$ is the metric of spacetime,
$\hom^{ij}$ is the metric-compatible connection, $\hnab\hg=0$, and
$\hT^i$ and $\hR^{ij}$ are the torsion and the RC curvature of
spacetime, respectively. In our convention, hatted variables are 3D
objects. Clearly, general features of PGT make it dynamically quite
different from Riemannian theories, such as, for instance,
topologically massive gravity \cite{x14,x15} or the
Bergshoeff--Hohm--Townsend gravity \cite{x16}.

In 3D, to any antisymmetric form $\hat X^{ij}$ there corresponds its
Lie dual form $\hat X_k$, defined by $\hat X^{ij}=-\ve^{ijk}\hat X_k$.
Replacing $\hom^{ij},\hR^{ij}$ with their Lie duals $\hom^i,\hR^i$, we
have:
\be
\hT^i=d\he^i+\ve^i{}_{jk}\hom^j\wedge\he^k\, ,\qquad
\hR^i=d\hom^i+\frac{1}{2}\ve^i{}_{jk}\hom^j\wedge\hom^k\,
\ee

In local coordinates $x^\m$, we can write $\he^i=\he^i{_\m}dx^\m$,
$\hom^i=\hom^i{_\m}dx^\m$, and the action of local Poincar\'e
transformations on the basic dynamical variables reads:
\bea
&&\d_0\he^i{_\mu}=-\ve^{ijk}\he_{j\mu}\hth_k-(\pd_\m\hxi^\l)\he^i{_\l}
   -\hxi^\l\pd_\l\he^i{_\m}\, ,                            \nn\\
&&\d_0\hom^i{_\mu}=-\hnab_\m\hth^i-(\pd_\m\hxi^\l)\hom^i{_\l}
                   -\hxi^\l\pd_\l\hom^i{_\m}\, .
\eea
Here, $\d_0$ is the form variation of a field, the parameters $\hth^i$
and $\hxi^\m$ describe local Lorentz transformations and local
translations, respectively, and
$\hnab_\m\hth^i=\pd_\m\hth^i+\ve^i{}_{jk}\hom^j{_\m}\hth^k$.

Specific features of the RC geometry in 2D are described in Appendix \ref{geometry}.

\subsection{Restricting the local Poincar\'e symmetry}

In order to study the holographic structure of 3D gravity with torsion,
we assume that spacetime $M$ is a 3D manifold with a boundary $\pd M$
at spatial infinity; more precisely, $M$ is asymptotically
diffeomorphic to $R\times\pd M$. The gravitational content of $M$
implies that its geometric structure is of the RC type, whereas its
dynamics is determined by choosing an action integral, which produces
the field equations. Given the field equations, the asymptotic behavior
of $M$ is controlled by the asymptotic conditions. In the asymptotic
region, $M$ can be suitably parametrized by the local coordinates
$x^\m=(\r,x^\a)$, where $\r$ is a radial coordinate, such that $\r=0$
on $\pd M$. The asymptotic conditions are formulated as certain
conditions on the gravitational variables $\he^i$ and $\hom^i$ near the
boundary at $\r=0$.

The (asymptotic) radial foliation of $M$ is an analog of the
temporal foliation in the standard canonical formalism, with
time line replaced by the radial line; early ideas on dynamical
evolutions along spatial directions can be found in \cite{x17}.
In this framework, Poincar\'e gauge invariance implies that
$\he^i{_\r}$ and $\hom^i{_\r}$  are unphysical variables, so
that their values can be fixed by suitable gauge conditions.
Although gauge conditions have no influence on the physical
content in the bulk, the boundary dynamics is very sensitive to
their form. Based on the experience with the Mielke-Baekler (MB)
topological model of 3D gravity with torsion \cite{x8,x11}, we
impose the following six \emph{gauge conditions}:
\bsubeq\lab{2.3}
\bea
&&\he^i{_\r}=(\he^1{_\r},\he^a{_\r})
            =\left(\frac{\ell}{\r},0\right)\,,             \lab{2.3a}\\
&&\hom^i{_\r}=(\hom^1{_\r},\hom^a{_\r})
             =\left(\frac{p\ell}{2\r},0\right)\, ,         \lab{2.3b}
\eea
\esubeq
which break the Lorentz and the translational gauge invariance; $\ell$
is the AdS radius. As we shall see below, the parameter $p$ controls
the strength of both the torsion and the curvature on $M$. Next, we
impose an \emph{extra condition}:
\be
\he^1{_\a}=0\, ,                                           \lab{2.4}
\ee
which is equivalent to $\he_i{^\r}=0$ and is known as the ``radial
gauge'' (an analog of the standard ``time gauge''). Geometrically, it
ensures that the radial direction coincides with the normal to $\pd M$,
which greatly simplifies the calculations. In particular, the matrix
representation of $\he^i{_\m}$ becomes block diagonal. Finally,
combining \eq{2.4} with a suitable ansatz for $\he^a{_\a}$ and
$\hom^i{_\a}$, we can write:
\bsubeq\lab{2.5}
\bea
&&\he^i{_\a}=(\he^1{_\a},\he^a{_\a})
            =\left(0,\frac{1}{\r}e^a{_\a}\right)\, ,       \lab{2.5a}\\
&&\hom^i{_\a}=(\hom^1{_\a},\hom^a{_\a})
             =\left(\om_\a,\frac{1}{\r}k^a{_\a}\right)\, , \lab{2.5b}
\eea
\esubeq
where $e^a{_\a}(\r,x),\om_\a(\r,x)$ and $k^a{_\a}(\r,x)$ are assumed to
be finite and differentiable functions of $\r$ at $\r=0$, such
that, near the boundary, they have the form
\bea
&&e^a{_\a}(\r,x)=\bare^a{_\a}(x)+\cO(\r)\, ,                \nn\\
&&\om_\a(\rho,x)=\bom_\a(x)+\cO(\r)\, ,                     \lab{2.6}
\eea
and similarly for $\k^a{_\a}(\r,x)$. Here, $\cO(\r)$ tends to zero when
$\r\to 0$, a bar over $e^a{_\a}$ denotes the value of $e^a{_\a}$ at the
boundary $\r=0$, and similarly for $\bar\om_\a$. Note, in particular,
that the conditions \eq{2.6} allow the presence of $\r^n\ln\r$ terms
for $n>0$, but not for the leading term $n=0$. The inverse of
$\he^i{_\m}$ has the form
\bea
&&\he_i{^\r}=(\he_1{^\r},\he_a{^\r})
            =\left(\frac{\r}{\ell},0\right)\, ,            \nn\\
&&\he_i{^\a}=(\he_1{^\a},\he_a{^\a})=(0,\r e_a{^\a})\, .   \lab{2.7}
\eea
The geometric interpretation of $e^a{_\a},\om_\a$ and $k^a{_\a}$ will
be given in the next subsection.

Based on these conditions, we will investigate the holographic
structure of 3D gra\-vi\-ty with torsion, assuming \emph{the absence of
matter}. In particular, we shall study two complementary dynamical
situations, described by
\bitem
\item[(a)] MB model of topological 3D gravity with
torsion, and\vspace{-9pt}
\item[(b)] general (parity-preserving) 3D gravity with propagating
torsion.
\eitem

For later convenience, we note that the metric defined by \eq{2.3} and
\eq{2.5},
\be
ds^2=\hat g_{\m\n}dx^\m dx^\n=-\frac{\ell^2d\r^2}{\r^2}
     +\frac{1}{\r^2}g_{\a\b}dx^\a dx^\b \, ,               \nn
\ee
where $g_{\a\b}:=e^a{_\a}e^b{_\b}\eta_{ab}$ is regular at $\r=0$
and takes the usual Fefferman--Graham form \cite{x18}. For
$\r=0$, the full metric has a pole of order two, which is typical for
asymptotically AdS spaces, and directly related to the pole of order
one in the triad field \eq{2.5a}.

In the rest of the paper, we use the units in which the AdS radius
is $\ell=1$.

\prg{Comment on \eq{2.6}.} Any assumption on the asymptotic form of
dynamical variables restricts the set of possible solutions of the
field equations. In general, depending on the model-dependent dynamical
features, expansions of the fields in \eq{2.6}  could contain
logarithmic terms or power series of different order. However, having
in mind that the holographic structure of the general 3D gravity model
(b) has not been studied before, our intention is not to make the most
general holographic analysis, which would be technically extremely
complex, but to identify its basic holographic features. Furthermore,
since both models (a) and (b) possess asymptotically AdS black hole
solution \cite{x12}, it is quite natural to expect that those features
can be successfully revealed by focusing on the AdS asymptotic sector
of the Brown--Henneaux type \cite{x19,x11}.

To be more specific, let us mention that certain holographic aspects of
the MB model in the Chern-Simons formulation have been studied earlier
by Klemm and Tagliabue \cite{x8}. Their results strongly suggest that,
in the MB model, our assumption \eq{2.6} should be restricted to the
following form:
\bea
&&\he^a{_\a}(\r,x)=\bare^a{_\a}+\r^2s^a{_\a}+\cO_4\, ,     \nn\\
&&\hom_\a(\rho,x)=\bom_\a+\cO_2\, ,                        \lab{2.8}
\eea
where $\cO_n$ is a term that tends to zero as $\r^n$ or faster, when
$\r\to 0$. Moreover, we expect the same sector to be of prime interest
for the holographic structure of the general 3D gravity model (b). As
we shall see, the results obtained in sections 4 and 5 justify our
expectations. In this section, however, we continue using only
\eq{2.6}.

\subsection{Residual gauge symmetries}

A field theory is defined by both the field equations and the
asymptotic (boundary) conditions. The concept of asymptotic symmetry is
of fundamental importance for understanding basic aspects of the
boundary dynamics. Since the conditions \eq{2.3}, \eq{2.5} and \eq{2.6}
control the form of dynamical variables in the asymptotic region near
$\r=0$, they have a decisive influence on the asymptotic symmetry. The
asymptotic symmetry is defined by a subset of gauge transformations
that leaves the asymptotic conditions invariant. Thus, the parameters
of the asymptotic (or residual) gauge transformations are defined by
the consistency requirements
\bea
&&-\ve^{ijk}\he_{j\mu}\hth_k-(\pd_\m\hxi^\l)\he^i{_\l}
          -\hxi^\l\pd_\l\he^i{_\m}=0\, ,                   \nn\\
&&-\nab_\m\hth^i-(\pd_\m\hxi^\l)\hom^i{_\l}
          -\hxi^\l\pd_\l\hom^i{_\m}=0\, ,                  \nn
\eea
where $\he^i{_\m}$ and $\hom^i{_\m}$ are taken to satisfy \eq{2.3} and
\eq{2.5}.

Starting with these conditions, we first find the restrictions stemming
from the invariance of $\he^1{_\r},\he^a{_\r}, \he^1{_\a}$, and
$\hom^1{_\r}$, respectively:
\bsubeq\lab{2.9}
\bea
&&\hxi^\r=\r f(x)\, ,                                      \nn\\
&&\pd_\r\hxi^\a=\r g^{\a\b}\pd_\b f\,,                     \nn\\
&&\hth^a=\r\ve^{ab}e_b{^\a}\pd_\a f\, ,                    \nn\\
&&\pd_\r\hth^1=-\r\om^\a\pd_\a f\, .                       \lab{2.9a}
\eea
There relations give rise to the following radial radial expansion of
the local parameters:
\bea
&&\hxi^\r=\r f(x)\, ,                                      \nn\\
&&\hxi^\a=\xi^\a(x)+\frac{1}{2}\r^2\bg^{\a\b}\pd_\b f
                   +\r^2\cO(\r)\, ,                        \nn\\
&&\hth^a=\r\ve^{ab}\bare_b{^\a}\pd_\a f+\r\cO(\r)\, ,      \nn\\
&&\hth^1=\th(x)-\frac{\r^2}{2}\bom^\a\pd_\a f+\r\cO(\r^2)\,.\lab{2.9b}
\eea
\esubeq
Thus, the residual symmetry is expressed in terms of the \emph{four
boundary parameters}: $\xi^\a(x),\th(x)$ and $f(x)$.

In the next step, we find the restrictions produced by the invariance
of $\hom^a{_\r}$ and $\hom^a{_\a}$, respectively:
\bea
&&\left[\left(\ve^{ab}-\frac{p}{2}\eta^{ab}
         +k^{ab}\right)e_b{^\b}
  +\r\ve^{ab}(\pd_\r e_b{^\b})\right]\pd_\b f=0\, ,        \nn\\
&&\d_0 k^a{_\a}=[-\ve^{ab}k_{b\a}\th-(\pd_\a\xi^\b)k^a{_\b}
  -\xi^\b\pd_\b k^a{_\a}] +f k^a{_\a}+\cO(\r)\, .          \lab{2.10}
\eea
Assuming that $f(x)$ is an arbitrary function on $\pd M$, we have
$\pd_\b f\ne 0$, and the first relation defines $k^a{_\a}$ in terms of
the $e^a{_\a}$:
\be
k^{ab}=\frac{p}{2}\eta^{ab}-\ve^{ab}
       -\r\ve^{ac}e_c{^\b}\pd_\r e^b{_\b}\, ,              \lab{2.11}
\ee
where $k^{ab}=k^a{_\a} e_b{^\a}$. The second relation in \eq{2.10}
defines the transformation law for $k^a{_\a}$; it shows that, at the
boundary, $k^a{_\a}$ is a tensorial object with respect to local
Poincar\'e transformations combined with dilatations. As shown in
Appendix \ref{geometry}, $K_{ab}=\ve_{cb}k^c{_a}$ is the extrinsic curvature of $\pd
M$.

Finally, we wish to examine the implications of the invariance
conditions for $\he^a{_\a}$ and $\hom^1{_\a}$. Using \eq{2.5}, these
condition yield, in the lowest order of the radial expansion, the
following transformation rules for the boundary fields:
\bea
&&\d_0\bare^a{_\a}=\d_P\bare^a{_\a}+f\bare^a{_\a}\, ,      \nn\\
&&\d_0\bom_\a=\d_P\bom_\a
              +\ve_{ab}\bare^a{_\a}\bare^{b\b}\pd_\b f\, , \lab{2.12}
\eea
where $\d_P\bare^a{_\a}$ and $\d_P\bom_\a$ are the local Poincar\'e
transformations in 2D:
\bea
&&\d_P\bare^a{_\a}=-\ve^a{_c}\th\bare^c{_\a}
  -(\pd_\a\xi^\b)\bare^a{_\b}-\xi\cdot\pd\bare^a{_\a}\, ,  \nn\\
&&\d_P\bom_\a=-\pd_\a\th
 -(\pd_\a\xi^\b)\bom_\b-\xi\cdot\pd\bom_\a\, ,             \lab{2.13}
\eea
and $f$ defines local dilatations. Thus, we conclude the following:
\bitem
\item[$-$] The residual symmetry transformations \eq{2.12} belong to the
Weyl group of local Poincar\'e transformations plus dilatations,
whereas $\bare^a{_\a}$ and $\bom_\a$ are recoginzed as the vielbein and
the spin connection of the boundary RC geometry.
\eitem

The transformation rule for $\bare^a{_\a}$ can be used to calculate how
the residual symmetries act on the boundary metric
$\bg_{\a\b}=\eta_{ab}\bare^a{_\a}\bare^b{_\b}$. Restricting our
attention to dilatations ($f\ne 0$), we obtain
$\d_f\bar{g}_{\a\b}=2f\,\bar{g}_{\a\b}$.
For more details, see Appendix \ref{residual}.

The results obtained in this subsection are based only on the adopted
holographic conditions \eq{2.3}, \eq{2.5} and \eq{2.6}. We consider
them as being kinematical, in the sense that they are not influenced by
the dynamical arguments encoded in \eq{2.8}. Another useful set of
kinematical relations is found by calculating the expressions for the
torsion and the curvature tensors, based on \eq{2.3}, \eq{2.5} and
\eq{2.6}. As shown in Appendix \ref{R and T}, the result is of the form
\bsubeq\lab{2.14}
\be
\hT_{ijk}=p\ve_{ijk}+\cO(\r)\, ,\qquad
\hR_{ijk}=q\ve_{ijk}+\cO(\r)\, ,                           \lab{2.14a}
\ee
where
\be
q:=\frac{p^2}{4}-1\, .                                     \lab{2.14b}
\ee
\esubeq
Thus, to lowest order in $\r$, the parameter $p$ defines both the
torsion and the curvature of spacetime.

In sections 4 and 5, we shall combine these results with \eq{2.8}
to study the specific dynamical models.

\section{Noether--Ward identities}
\setcounter{equation}{0}

It is clear from the previous discussion that the residual gauge
symmetries \eq{2.9} are also kinematical. They are \emph{maximal} gauge
symmetries that we can expect to find on the boundary. Indeed, after
choosing an action integral, the corresponding field equations may
impose additional restrictions on these symmetries. In this section, we
shall study the gravitational Noether identities (also called
generalized conservation laws) induced by the maximal gauge symmetries
\eq{2.9}, and interpret them as the corresponding Ward identities of
the boundary CFT.

To make these ideas more precise, consider a 3D gravitational system
without matter in an asymptotically AdS spacetime, with solutions
characterized by independent boundary values of $e^a{_\a}$ and
$\om_\a$. The quasilocal energy-momentum and spin currents of the
system are calculated by varying the action with respect to the
boundary values of $e^a{_\a}$ and $\om_\a$. The variation produces a
bulk term, which is proportional to the field equations, and a boundary
term. The on-shell value of the gravitational action, suitably
\emph{renormalized}, is given as a \emph{finite} 2D functional $I_{\rm
ren}[e,\om]$ on $\pd M$. Next, consider a set of quantum fields $\phi$
on $\pd M$, coupled to the external gravitational fields (sources)
$e^a{_\a}$ and $\om_\a$, and described by an action integral
$I[\phi;e,\om]$. The corresponding effective action $W[e,\om]$ is
defined by the functional average over $\phi$:
\bsubeq\lab{3.1}
\be
e^{iW[e,\om]}=\int_{\pd M}D\phi e^{iI[\phi;e,\om]}\, .     \lab{3.1a}
\ee
In the semiclassical approximation, the AdS/CFT correspondence can
be expressed by identifying the effective action with $I_\ren[e,\om]$:
\be
W[e,\om]=I_\ren[e,\om]\, .                                 \lab{3.1b}
\ee
\esubeq
Using this identification, we can calculate the gravitational
\emph{Noether identities} for $I_\ren[e,\om]$ and identify them as the
\emph{Ward identities} for the 1-point functions derived from
$W[e,\om]$, provided the functional measure is invariant under the
residual gauge symmetries.

We consider gravity theories whose Lagrangians are at most
quadratic in the first derivatives of the spin connection and the
vielbein. The corresponding field equations are obtained integrating by
parts, such that the surface term,
\begin{equation}
\delta I_{\mathrm{on-shell}}=\int d^{2}x\,\left( P_{i}^{\nu }\,\delta \hat{e}
_{\,\, \nu }^{i}+Q_{i}^{\nu }\,\delta \hat{\omega}_{\,\, \nu }^{i}\right)
\,,
\end{equation}
does not contain derivatives of the variations of the fields.

The gauge choice \eq{2.3}--\eq{2.5}, when used in the above
formula, produces a surface term expressed in terms of the
boundary quantities
\begin{equation}
\label{var_I}
\delta I_{\mathrm{on-shell}}=\int d^{2}x\,\left( p_{i}^{\alpha }\delta
e_{\,\, \alpha }^{a}+q^{\alpha }\delta\omega _{\,\, \alpha}+
\tilde{q}_{a}^{\alpha }\delta k_{\,\, \alpha }^{a}\right) \,.
\end{equation}
It is clear that the PGT formulation of gravity also allows to
impose boundary conditions different than keeping the
vielbein and spin connection fixed at
the conformal boundary. However, a theory with boundary conditions
other than a Dirichlet one does not lead itself to a holographic
description in the usual AdS/CFT framework.

In fact, in the metric formalism, the last term
in (\ref{var_I}) is related to the variation of the extrinsic curvature
that is usually traded off for the variation of metric by
a Gibbons-Hawking-type term. When a Gibbons-Hawking-type
term cannot be constructed for a given theory,
the only way out is to consider that the extrinsic curvature and the metric are related
asymptotically.

The fact that the leading-order in the expansion of the extrinsic
curvature is the same as the leading-order of the boundary metric for
Riemannian AdS spacetimes suggests that there is an asymptotic relation between the
extrinsic curvature and the vielbein in theories with torsion; such a relation in PGT is given by \eq{2.11}.
Note that, as showed in Appendix \ref{geometry}, only the symmetric part of the extrinsic curvature is Riemannian,
and the antisymmetric one explicitly depends on torsion. Once appropriate counterterms are added, the variation of the
renormalized PGT action can be written as
\begin{equation}
\d~I_\ren= -\int_{\pd M}d^2 x
   \left(\t^\a{_a}\d e^a{_\a}+\s^\a\d \om_\a\right)\, ,
\end{equation}
whereby the standard duality between gravity and a boundary
CFT is recovered.

The form of the expected Noether identities is based on the residual
symmetry transformations \eq{2.12} and \eq{2.13}. Quite generally, the
invariance of the renormalized action under these transformations can
be written in the form
\bsubeq
\be
\d~I_\ren= -\int_{\pd M}d^2 x
   \left(\t^\a{_a}\d_0e^a{_\a}+\s^\a\d_0\om_\a\right)=0\, ,\lab{3.2a}
\ee
where
\be
\t^\a{_a}:=-\frac{\d I_\ren}{\d e^a{_\a}}\, ,\qquad
  \s^\a:=-\frac{\d I_\ren}{\d\om_\a}\, ,                   \lab{3.2b}
\ee
\esubeq
are the energy-momentum and spin currents (tensor densities) of
our dynamical system.

Restricting our attention first to the local translations (with
parameters $\xi^\a$) and then to the local Lorentz transformations
(with parameter $\th$), we arrive at the corresponding Noether
identities:
\bsubeq\lab{3.3}
\bea
&&e^a{_\b}\nabla_\a\t^\a{_a}=\t^\a{_a}T^a{}_{\b\a}+\s^\a F_{\b\a}
  -\om_\b(\nab_\a\s^\a+\ve^{ab}\t_{ab})\, ,                \lab{3.3a}\\
&&\nab_\b\s^\b=-\ve^{ab}\t_{ab}\, ,                        \lab{3.3b}
\eea
which are also known as the generalized conservation laws of
$\t^\a{_a}$ and $\s^\b$. Note that if the second Noether identity
\eq{3.3b} is fulfilled, the last term in \eq{3.3a} can be omitted.
Similarly, the invariance of $I_\ren[e^a{_\a},\om_\a]$ under
dilatations leads to
\be
\t-\nabla_\b\bigl(\ve_{ab}\s^ae^{b\b}\bigr)=0\, ,         \lab{3.3c}
\ee
\esubeq
where $\t:=\t^a{_a}$ is the trace of the energy momentum tensor.

Although the gravitational dynamics in the bulk is described by a
RC geometry, with $\hom^i{_\m}$ and $\he^i{_\m}$ as independent fields,
it may happen that some solutions on the boundary are Riemannian, that
is, characterized by a vanishing torsion, $T_{abc}=0$. For such
solutions, the boundary connection $\om_\a$ is no longer independent of
the vielbein $e^a{_\a}$. Nevertheless, as we are going to show, the
Noether--Ward identities are still of the form \eq{3.3}, but now,
$\om_\a$ takes on the Riemannian value $\tom_\a$. In a way, this might
have been expected, since the transformation properties of $\tom_\a$
are the same as those of $\om_\a$, and these properties play a crucial
role in defining the boundary symmetry.

When the boundary torsion vanishes, the connection takes the Riemannian
form \eq{A.3}. However, we find it more convenient to use an equivalent
but more compact expression:
\be
\tom_\a=-\ve_{ab}\ve^{\g\d}\ve_{\a\b}e^{a\b}\pd_\g e^b{_\d}\,.\lab{3.4}
\ee
Now, starting from the Riemannian renormalized action $\tilde
I_\ren=I_\ren[e^a{_\a},\tom_\a]$, we find that the related spin
current $\S^\a:=-\d\tilde I_\ren/\d\om_\a$ vanishes, whereas the
energy-momentum current $\Th^\a{_a}:=-{\d\tilde I_\ren}/{\d e^a{_\a}}$
has an additional contribution stemming from the last term in
\eq{3.2a}:
\be
\Th^\a{_a}=\tilde\t^\a{_a}
   -\tilde\nab_\b\left(\ve^{\a\b}e^{-1}\tilde\s_a\right)\, .
\ee
Here, $\tilde X$ denotes the Riemannian limit of a RC object $X$; in
particular, $\tilde\nab_\a f_a=\pd_\a f_a -\ve_{ac}\tom_\a f^c$. Then,
the Noether identities for the action $\tilde I_\ren$ are found to be:
\bsubeq\lab{3.6}
\bea
&&e^a{_\b}\tilde\nab_\a\Th^\a{_a}+\tom_\b\ve^{ab}\Th_{ab}=0\,,\lab{3.6a}\\
&&\ve^{ab}\Th_{ab}=0\, ,                                     \\
&&\Th=0\, .
\eea
\esubeq
Since $\Th^\a{_a}$ is a tensor density, the first relation, which is a
condition for diffeomorphism invariance, is seen to coincide with the
condition (4.10) in Klemm et al. \cite{x8}. When the Lorentz invariance
is satisfied, \eq{3.6a} reduces to the usual form
$D_\a(e^{-1}\Th^\a{_\b})=0$, where $D_\a$ is the Riemannian covariant
derivative. The remaining two relations are the standard Riemannian
conditions for the Lorentz and Weyl invariance, respectively. Using
$T_{abc}=0$, as well as the identity
$\ve^{\a\b}\tilde\nab_\a\tilde\nabla_\b f_a
  =-\frac{1}{2}\ve^{\a\b}\tilde F_{\a\b}\ve_{ab}f^b$,
one can transform \eq{3.6} into
\bsubeq\lab{3.7}
\bea
&&e^a{_\b}\tilde\nab_\a\tilde\t^\a{_a}-\tilde\s^\a F_{\b\a}
      +\tom_\b(\tilde\nab_\b\tilde\s^\b+\ve^{ab}\tilde\t_{ab})=0\,,\\
&&\ve^{ab}\tilde\t_{ab}+\tilde\nab_\b\tilde\s^\b=0\, ,     \\
&&\tilde\t-\tilde\nab_\b(\ve_{ab}\tilde\s^a e^{b\b})=0\, .
\eea
\esubeq
Hence, the Riemannian identities \eq{3.6} coincide with those obtained
from \eq{3.3} in the limit $T_{abc}\to 0$, as expected. This proves the
following theorem:
\bitem
\item[$-$] In the context of PGT, the form \eq{3.3} of Noether
identities can be used for both Riemann--Cartan and Riemannian boundary
geometries.
\eitem

According to the AdS/CFT correspondence, relations \eq{3.3} are
interpreted as the maximal set of Ward identities that can be found in
the boundary CFT. If the field equations happen to be incompatible with
the above symmetries, some of the Ward identities may be violated,
leading to the appearance of \emph{quantum anomalies}.

\section{Holography in topological 3D gravity with torsion}
\setcounter{equation}{0}

In this section, we analyze the validity of the Noether--Ward
identities  \eq{3.3}, in the MB model of topological 3D gravity
with torsion \cite{x9,x11}, described by the action
\be
I_\mb=\int\left(2a\he^i\hR_i-\frac{1}{3}\L_0\ve_{ijk}\he^i\he^j\he^k
      +\a_3L_{\rm CS}(\hom)+\a_4\he^i\hT_i\right)\, ,
\ee
where $L_{\rm CS}(\hom)=\hom_i d\hom^i
+\frac{1}{3}\ve_{ijk}\hom^i\hom^j\hom^k$ is the Chern--Simons
Lagrangian for the Lorentz connection,  $a=1/16\pi G$ is the
gravitational constant, $\L_0$ is a (bare) cosmological constant,
$\a_3,\a_4$ are dimensionful coupling constants, the wedge product
signs $\wedge$ are omitted for simplicity, and the matter
contribution is absent.

The vacuum field equations read
\bsubeq\lab{4.2}
\be
\hT_{ijk}=p\ve_{ijk}\, ,\qquad \hR_{ijk}=q\ve_{ijk}\, ,    \lab{4.2a}
\ee
where the parameters $p$ and $q$ are defined in terms of the coupling
constants $a,\L,\a_3,\a_4$. The spacetime described by these equations
is maximally symmetric, at least locally. Moreover, in the AdS sector,
the effective cosmological constant is negative,
\be
\L_{\rm eff}:=q-\frac{p^2}{4}=-1\, .                       \lab{4.2b}
\ee
\esubeq
By comparing these equations with \eq{2.14}, it follows that the
parameter $p$ from our ansatz should be identified with the parameter
$p$ in the MB model.

Our analysis is based on using the AdS asymptotic conditions \eq{2.8}.
For an interesting asymptotic correspondence between the MB model and
topologically massive gravity, see \cite{x20}.

\subsection{Analysis of the field equations}

The subset of the field equations \eq{4.2a} that describes the
\emph{radial evolution} of the system is given by $(ijk)=(11c),(a1c)$.
The first pair of equations takes the form
\bsubeq
\be
\hT_{11c}=0\, ,\qquad \hR_{11c}=0\, .
\ee
Using the expressions for $\hT_{ijk}$ and $\hR_{ijk}$ calculated in
Appendix \ref{F}, one finds that the first equation is identically satisfied,
whereas the second one implies that $\om_\a$ is the Lorentz connection
at the boundary,
\be
\om_\a=\om_\a(x)\, .
\ee
\esubeq
The second pair of equations reads:
\bsubeq\lab{4.4}
\be
\hT_{a1c}=-p\ve_{ac}\, ,\qquad \hR_{a1c}=-q\ve_{ac}\, .    \lab{4.4a}
\ee
After introducing the radial expansion \eq{2.6}, the first equation
in \eq{4.4a} yields that $s_{ab}$ is symmetric,
\be
\ve^{ab}s_{ab}=0\, .                                       \lab{4.4b}
\ee
This result simplifies the second equation in \eq{4.4a}; relying on
\eq{B.9}$_3$, the piece of the zeroth order in $\r$ implies that the
effective cosmological constant $\Leff$ is negative, see \eq{4.2b},
whereas the piece of order $\r^2$ leads to a finite radial expansion of
$e_{c\b}$:
\be
e_{c\b}=\bare_{c\b}+\r^2\bar s_{c\b}\, .                   \lab{4.4c}
\ee
\esubeq Such an expansion that terminates at $\r^2$ is a
generalization of the result known for GR in 3D; in higher
dimensions, the result holds when the Weyl tensor vanishes
\cite{x21}. As a simple consequence, the radial expansion of
$k^{ab}$ is also finite:
$$
k^{ab}=\frac{p}{2}\eta^{ab}-\ve^{ab}+2\r^2\ve^{ac}s^b{_c}\, .
$$

Using the above results, the nontrivial content of the remaining
$(1bc)$ and $(abc)$ field equations is expressed in terms of the
following \emph{radial contraints}:
\bsubeq\lab{4.5}
\bea
&&T_{abc}=0\, ,                                            \lab{4.5a}\\
&&R-4s^c{_c}=\cO_2\, ,                                     \lab{4.5b}\\
&&\nab_\a s_{b\b}-\nab_\b s_{b\a}=0\, .
\eea
\esubeq
In particular, we see that the boundary torsion vanishes.

\subsection{Counterterm and boundary currents}

Now, we introduce the boundary currents and verify their Noether--Ward
identities.

The variation of the MB action, calculated on shell, reduces to a
surface integral:
\be
\d I_\mb=\int_{\pd M}d^2 x\ve^{\a\b}\left(
  2a\he^i{_\a}\d\hom_{i\b}+\a_3\hom^i{_\a}\d\hom_{i\b}
  +\a_4\he^i{_\a}\d\he_{i\b}\right)\, .
\ee
Each of these three terms can be written in more details as:
\bea
&&2a\ve^{\a\b}\he^i{_\a}\d\hom_{i\b}
  =\frac{2a}{\r^2}\ve^{\a\b}\left[
    \frac{p}{2}e^b{_\a}\d e_{b\b}
    -\ve_{ab}e^a{_\a}\d e^b{_\b}
    -2\r^2\ve_{ab}s^a{_\a}\d e^b{_\b}\right]+\d\D_1\, ,    \nn\\
&&\a_3\ve^{\a\b}\hom^i{_\a}\d\hom_{i\b}
  =\frac{\a_3}{\r^2}\ve^{\a\b}
   \left[qe^b{_\a}\d e_{b\b}
   -2p\r^2\ve^{ab}s_{a\a}\d e_{b\b}
   +4\r^2s^b{_\a}\d e_{b\b}\right]                         \nn\\
&&\hspace{3cm}   -\a_3\ve^{\a\b}\om_\a\d\om_\b\, ,         \nn\\
&&\a_4\ve^{\a\b}\he^i{_\a}\d\he_{i\b}=
  \frac{\a_4}{\r^2}
          \ve^{\a\b}e^b{_\a}\d e_{b\b}\, ,                 \nn
\eea
where $\d\D_1$ is a total variation with
$$
\D_1:=4a\ve^{\a\b}\ve_{ab}s^a{_\a}e^b{_\b}=-4a\bare s^c{_c}\,,
$$
and $e:=\det(e^a{_\a})$. Then, the identity $ap+\a_3 q+\a_4=0$,
see Ref. \cite{x11}, implies that the sum of the first three
terms in the above expressions vanishes, whereupon the only
\emph{divergent} term in $\d I_\mb$ is also a total variation,
$\d\D_2$, with
$$
\D_2:=-\frac{a}{\r^2}\ve^{\a\b}\ve_{ab}e^a{_\a}e^b{_\b}
     =\frac{2a}{\r^2}\bare(1+\r^2 s^c{_c})\,.
$$

Since the boundary integral of $\D_1+\D_2$ appears in $\d I_\mb$ as a
total variation, it can be subtracted from $I_\mb$ to obtain an
\emph{improved} variational principle. The integral
\bsubeq\lab{4.7}
\be
I_\ct:=\int_{\pd M} d^2x(\D_1+\D_2)
      =2a\int_{\pd M}d^2x\bare \left(\frac{1}{\r^2}-s^c{_c}\right) \lab{4.7a}
\ee
is usually called the \emph{counterterm}. Before discussing its role in
the new variational principle, let us rewrite $I_\ct$ in an equivalent
form as
\be
I_\ct:=a\int_{\pd M}d^2x\te K\, ,
\ee
where $K$ is the trace of the extrinsic curvature \eq{A.5}, and
$\te^a{_\a}=\bare^a{_\a}/\r\,$ is the induced vielbein at the
boundary. The expression for $I_\ct$ is just one-half of the
Gibbons--Hawking term ($I_{\rm GH}$), the result that naturally
appears in the Chern--Simons formulation of GR in 3D, as
discussed by Ba\~nados and M\'endez \cite{x22}, and by Mi\v
skovi\'c and Olea \cite{x23} (for an interesting approach to
counterterms in higher dimensional gravity, see \cite{x24}). On
the other hand, using the field equation \eq{4.5b}, we can
express the finite piece of the counterterm, $s^c{_c}$, in terms
of the scalar curvature $R$, but since $R$ is a topological
invariant, its contribution to $I_\ct$ can be disregarded. Thus,
effectively, the counterterm can be written as a covariant
object, determined by a local function of $\te^a{_\a}$:
\be
I_\ct=2a\int d^2 x\te=I_{{\rm GH}}-2a\int d^2 x\te\,,
\ee
\esubeq
where the last term is the usual \emph{local
counterterm} of Balasubramanian and Kraus \cite{x25}, obtained
in the context of 3D GR. It is interesting to note that the
nonlinear Chern--Simons term in the MB action does not
contribute to the counterterm, in agreement with the analysis of
\cite{x7}.

Since the total variation $\d I_\ct$ is a divergent piece of $\d
I_\mb$, we are quite naturally led to introduce the \emph{renormalized}
(or, more precisely, the improved) MB action:
\be
I^\ren_\mb:=I_\mb-I_\ct\, ,
\ee
such that it has well-defined functional derivatives and produces
finite boundary currents, on-shell.

Note that, although the counterterm \eq{4.7a} ensures that the  variation of $I^\ren_\mb$ is
finite and differentiable, one can verify that the value of the renormalized action $I^\ren_\mb$ is logarithmically divergent.
Similarly as in GR, the logarithmic term is proportional to the Euler topological invariant $eR$, which is why it does not influence the variation of $I^\ren_\mb$.
The logarithmic terms, even though topological in three dimensions, are important to
be included, because the renormalized gravitational action is identified
with the free energy in the dual boundary field theory.

Finally, by noting that
\be
\d I^\ren_\mb=\int_{\pd M} d^2 x\ve^{\a\b}\left[
  -4\left(a+\frac{\a_3 p}{2}\right)\ve_{ab} s^a{_\a}\d e^b{_\b}
  +4\a_3 s_{b\a}\d e^b{_\b}-\a_3\om_\a\d\om_\b\right]\, ,
\ee
we can use \eq{3.2b} to obtain the energy-momentum and spin currents on
the boundary:
\bea
&&\t^\b{_b}
   =4\left(a+\frac{\a_3 p}{2}\right)\ve^{\a\b}\ve_{ab}s^a{_\a}
    -4\a_3\ve^{\a\b}s_{b\a}\, ,                            \nn\\
&&\s^\b=-\a_3\ve^{\b\a}\om_\a\, .
\eea

\subsection{Boundary symmetries and anomalies}

Now, we wish to check the expected Noether--Ward identities \eq{3.3}.

Using the radial constraints \eq{4.5}, we find the following on-shell
relations:
\be
\nab_\b\t^\b{_b}=0\, ,\qquad
\nab_\b\s^\b=-\frac{1}{2}\ve^{bc}\t_{bc}\, .               \lab{4.11}
\ee
Comparing with \eq{3.3b}, we see that the Lorentz invariance of the
effective 2D theory is violated, and the \emph{Lorentz anomaly} reads:
\be
A_{\rm L}:=\nab_\b\s^\b+\ve^{bc}\t_{bc}
    =\frac{1}{2}\ve^{bc}\t_{bc}=-\frac{1}{2}\a_3\bare R\,. \lab{4.12}
\ee
The coefficient $\a_3$, multiplying the topological (Euler)
density $\bare R$, is proportional to the difference of the
classical central charges $c^\mp$ of the Mielke--Baekler model
\cite{x11}:
$$
c^\mp=24\pi\left[a\ell+\a_3\left(\frac{p\ell}{2}\mp 1\right)\right]\, .
$$

Next, \eq{4.11}$_1$ implies that the translation invariance condition
\eq{3.3b} is reduced to the form $0=\s^\b F_{\a\b}+\om_\a
\nab_\b\s^\b$. Using the relations
$$
\nabla_\b\s^\b=\frac{1}{2}\a_3 eR\, ,\qquad
  \s^\b F_{\a\b}=-\frac{1}{2}\a_3\om_\a eR\, ,
$$
we conclude that local translations are a correct boundary symmetry.
Hence, there is \emph{no translational} anomaly:
\be
A_{\rm T}:=e^a{_\b}\nabla_\a\t^\a{_a}
  -\t^\a{_a}T^a{}_{\b\a}-\s^\b F_{\a\b}
  +\om_\a(\nab_\b\s^\b+\ve^{ab}\t_{ab})=0\, .              \lab{4.13}
\ee

Finally, in order to verify the Noether identity for dilatations
\eq{3.3c}, we use \eq{4.4b} and \eq{4.5b} to obtain
\bea
&&\t^c{_c}=-4\bar e\left(a+\frac{\a_3p}{2}\right)s^c{_c}
          =-\bar e\left(a+\frac{\a_3p}{2}\right)R\, ,      \nn\\
&&\nab_\b\left(\ve_{ab}\s^ae^{b\b}\right)
   =-\a_3\pd_\b(\bar e\om^\b)\, .
\eea
Thus, the dilatational Noether identity is violated, and the violation
is measured by a quantity which is usually called the \emph{conformal}
(or Weyl) \emph{anomaly}:
\be
A_C:=\t^c{_c}-\nab_\b\left(\ve_{ab}\s^ae^{b\b}\right)
   =-\left(a+\frac{\a_3p}{2}\right)\bare R
    +\a_3\pd_\b(\bar e\om^\b)\, .                          \lab{4.15}
\ee
Here, the coefficient of $\bare R$ is proportional to the sum of the
central charges.

In treating the boundary symmetries of the MB model, Klemm et
al. \cite{x8} followed a different approach, based on using the
Riemannian connection in the renormalized action. Nevertheless,
our results for anomalies coincide with theirs, in agreement
with the theorem proved in section 3. The full strength of this
theorem will be seen in the more interesting case of 3D gravity
with propagating torsion, where the complicated field equations
may lead to either vanishing or nonvanishing boundary torsion.
However, we will be able to derive the Noether--Ward identities
without recourse to the value of the boundary torsion.

\section{Holography in 3D gravity with propagating torsion}
\setcounter{equation}{0}

In this section, we analyze the holographic structure of 3D gravity
with propagating torsion, assuming parity invariance \cite{x12},
and using the AdS asymptotic conditions \eq{2.8}.

\subsection{Lagrangian and the field equations}

Assuming the absence of matter, dynamical content of 3D gravity
with propagating torsion is defined by the action integral
\be
I=\int d^3 x\,\he\cL_G\, ,                                 \lab{5.1}
\ee
where $\he=\det(\he^i{_\m})$, and the gravitational
Lagrangian $\cL_G$ is at most \emph{quadratic} in the torsion
and the curvature. Assuming parity invariance, the general form
of $\cL_G$ is given by \cite{x12}
\bsubeq\lab{5.2}
\be
\cL_G=-a\hR-2\L_0+\cL_{T^2}+\cL_{R^2}\, .                  \lab{5.2a}
\ee
The quadratic terms can be conveniently be written in the form
\bea
&&\cL_{T^2}=\frac{1}{4}\hT^{ijk}\cH_{ijk}\, ,\qquad
  \cH_{ijk}:=a_1\,{}^{(1)}\hT_{ijk}+a_2\,{}^{(2)}\hT_{ijk}
                           +a_3\,{}^{(3)}\hT_{ijk}\, ,     \nn\\
&&\cL_{R^2}=\frac{1}{8}\hR^{ijkl}\cH_{ijkl}\, ,\qquad
  \cH_{ijkl}:=b_4\,{}^{(1)}\hR_{ijkl}+b_5\,{}^{(2)}\hR_{ijkl}
                           +b_6\,{}^{(3)}\hR_{ijkl}\, ,
\eea
where we introduced the covariant field momenta $\cH_{ijk}$ and
$\cH_{ijkl}$, which are \emph{linear} in the irreducible components of
the torsion, ${}^{(n)}\hT_{ijk}$, and the curvature,
${}^{(n)}\hR_{ijkl}$. An equivalent form of these two terms, which is
more convenient for practical calculations, is given by:
\bea
&&\cH_{ijk}=4(\a_1\hT_{ijk}+\a_2\hT_{[kj]i}+\a_3\hT_{ijk})\,,\nn\\
&&\cL_{R^2}=\hR^{ij}\cH_{ij}\, ,\qquad
  \cH_{ij}=\b_1\hR_{ij}+\b_2\hR_{ji}+\b_3\eta_{ij}\hR\, \label{RR}.
\eea
The expression for $\cL_{R^2}$ is obtained using the fact that
the Weyl tensor identically vanishes in 3D, and the new coupling
constants $(\a_k,\b_k)$ are expressed in terms of the
$(a_k,b_k)$ as \cite{x12}
\bea
&&\a_1=\frac{1}{6}(2a_1+a_3)\, ,\qquad \a_2=\frac{1}{3}(a_1-a_3)\,,
  \qquad \a_3=\frac{1}{2}(a_2-a_1)\, ,                     \nn\\
&&\b_1=\frac{1}{2}(b_4+b_5)\, ,\qquad\b_2=\frac{1}{2}(b_4-b_5)\,,
  \qquad \b_3=\frac{1}{12}(b_6-4b_4)\, .                   \nn
\eea
\esubeq

The variation of the action \eq{5.1} with respect to
$\he^i{_\m}$ and $\hom^{ij}{_\m}~(=-\ve^{ij}\hom_\m)$ produces
two gravitational field equations, displayed in equations (2.13)
of Ref. \cite{x12}. Without matter contribution, these
equations, transformed to the local Lorentz basis, take the
form: \bsubeq\lab{5.3}
\bea
&&\nab^m\cH_{imj}
  +\frac{1}{2}\cH_i{}^{mn}(-T_{jmn}+2\eta_{jm}V_n)
  -t_{ij}=0\, ,                                             \lab{5.3a}\\
&&t_{ij}:=\eta_{ij}\cL_G-T^{mn}{_i}\cH_{mnj}+2a\hR_{ji}
  -2(\hR^n{_i}\cH_{nj}-\hR_j{}^{nm}{_i}\cH_{nm})\, ,        \nn
\eea
where $t_{ij}$ is the energy-momentum tensor of gravity, and
\be
2aT_{kij}+2T^m{}_{ij}(\cH_{mk}-\eta_{mk}\cH)
  +4\nab_{[i}(\cH_{j]k}-\eta_{j]k}\cH)
  +\ve_{ijn}\ve^{mr}{_k}\cH_{mr}{^n}=0\, ,                  \lab{5.3b}
\ee
\esubeq
with $\cH=\cH^k{_k}$.

In the near-boundary expansion, the leading order of the field
equations \eq{5.3}, corresponding to $\r=0$, reduces to the
following relations involving the coupling constants:
\bsubeq\lab{5.4}
\bea
&&p(a+qb_6+2a_3)=0\, ,                                     \lab{5.4a}\\
&&aq-\L_0+\frac{1}{2}p^2a_3-\frac{1}{2}q^2b_6=0\, .        \lab{5.4b}
\eea
\esubeq As shown in \cite{x12}, these relations ensure that the
AdS configuration, as well as the black hole with torsion, are
solutions of the present theory. However, quadratic equations
\eq{5.4} allow to have two different solution for the effective
cosmological constant $\Leff=p-q^2/4$, and consequently, two
different AdS vacua. For a particular choice of parameters
($p=0, a-b_6q=0$), the two vacua coincide \cite{x12}. For an
analysis of this situation in the Bergshoeff--Hohm--Townsend
gravity, see Refs. \cite{x26,x27}.

\subsection{Equations of motion}

In this section, we discuss the consistency of the near-boundary
analysis of the field equations \eq{5.3}, given in Appendix \ref{expansion}, with the
holographic description of the asymptotic theory.

The leading order of the field equations is given by Eqs. \eq{5.4}.
These equations constrain the coupling constants and, therefore,
restrict the form of the allowed gravity actions.

Equations linear in $\r$ are given by the algebraic system (\ref{1a}),
(\ref{a1}), (\ref{11b}) and (\ref{cab}) for the vector $\bar{V}_a=\bar{T}^b{_{ba}}$,
which defines the complete torsion tensor in 2D (Appendix \ref{geometry}). These
equations allow not only Riemann--Cartan but also Riemann boundary
geometries. However, thanks to the theorem proved in section 3, we can
study the Noether--Ward identities in these two cases quite generally,
without making an explicit distinction between them.

The order $\r^2$ of the field equations is given in (\ref{11}--\ref{a1b}) and
(\ref{1ab}). These are algebraic equations in the tensor $s_{ab}$, which is
related to the extrinsic curvature $K_{ab}$ (Appendix \ref{expansion}). More
precisely, these equations determine the antisymmetric part
$\ve^{ab}s_{ab}$ and the trace $s^c{_c}$ as local expressions of the
boundary curvature and torsion. In particular, for the vanishing
torsion we have $\ve^{ab}s_{ab}=0$ and $s^c{_c}=\frac{1}{4}R$, as
in the MB model.

Here, in contrast to the MB model, the radial expansion goes beyond
$\r^2$, but the cubic and higher order terms do not affect our
results in the $\r \rightarrow 0$ limit.

Let us emphasize that, in our near-boundary analysis, we were not able
to determine the symmetric traceless part $s'_{ab}$ of $s_{ab}$. We can
understand this situation by noting that $s'_{ab}$ is a \emph{nonlocal}
function that requires a global solution. Such nonlocal terms are parts
of the (nonlocal) 1-point functions of the boundary CFT. On the other
hand, physical objects, such as the conformal anomaly, are always
local. This is a general feature of the boundary currents in an
effective theory.

In the next section, we calculate the boundary currents of the
effective CFT.

\subsection{Boundary currents}

In the absence of matter, the variation of the (gravitatonal)
action, evaluated \emph{on-shell}, takes the form
\begin{equation}
\d I_{\rm on-shell}=\int d^3x\,\pd _{\mu}
\left\{ 2\ve^{\m\n\l}\hat{e}^k{_\l}
\left[ \d\hat{e}^i{_\nu}\ve^{jm}{_k}\mathcal{H}_{ijm}
+\d\hat{\omega}^i{_\nu}\,\left( a\,\eta _{ik}+\mathcal{H}_{ki}
-\eta _{ki}\mathcal{H}\right) \right] \right\} \,.
\end{equation}
After expressing $\d I_{\rm on-shell}$ as a boundary integral, we will
use the field equations to find the renormalized 2D action. Then, in
accordance with \eq{3.2b}, we will identify the energy-momentum and the
spin boundary currents as the objects (1-point functions) coupled to
the sources $\bare^a{_\a}$ and $\bom_\a$ in the boundary CFT. To do
that, we write the action corresponding to the Lagrangian \eq{5.2a} as
\be
I=I_{EC}+I_{\L_0}+I_{T^2}+I_{R^2}\, .
\ee

The variation of the term $I_{EC}$, linear in the scalar curvature, is
known from the MB model:
\bsubeq
\bea
\d I_{EC}&=&\frac{ap}{\r^2}\int_{\pd\cM}d^2x
  \ve^{\a\b}e^a{_\a}\d e_{a\b}
  -4a\int_{\pd\cM}d^2x\ve^{\a\b}\ve^{ab}s_{\a a}\d e_{b\b} \nn\\
  &&+\d\int_{\pd M}d^2x(\D_1+\D_2)\, ,
\eea
where the total variation contains two pieces, one finite and the
other divergent:
\bea
\D_1&:=&4a\ve^{\a\b}\ve_{ab}s^a{_\a}e^b{_\b}=-4aes^c{_c}\, ,\nn\\
\D_2&:=&-\frac{a}{\r^2}\ve^{\a\b}\ve_{ab}e^a{_\a}e^b{_\b}
       =\frac{2a}{\r^2}\bare(1+\r^2 s^c{_c})\, .           \nn
\eea
\esubeq

The variation of the cosmological term does not contribute to the
boundary integrals. Next, we vary the term quadratic in torsion:
\be
\d I_{T^2}=\frac{2a_3p}{\r^2}\int_{\pd\cM}d^2x
           \ve^{\a\b}e^a{_\a}\d e_{a\b}
    +\frac{2a_3}{\r^2}\int_{\pd\cM}d^2x
          (\hcA-p)\ve^{\a\b}e^a{_\a}\d e_{a\b}\, .
\ee
Note that the second piece, containing the axial torsion, is a
finite 2D integral.

Finally, the variation of the term quadratic in curvature yields:
\bea
\d I_{R^2}&=&2\int_{\cM}d^3 x\ve^{\m\n\s}
    \left(\cH_{\s i}-b_{i\s}\cH\right)\pd_\m\d\hom^i{_\n}  \nn\\
&=&2\int_{\pd\cM}d^2x\ve^{\a\b}\left[\cH_{\a 1}\d\om_\b
   +\left(\cH_{ca}-\eta_{ac}\cH\right)
          \frac{1}{\r^2}e^c{_\a}\d k^a{_\b}\right]\, ,     \lab{5.8}
\eea
where
\be
\cH_{ca}-\eta_{ca}\cH=\eta_{ca}b_6q
  +\eta_{ca}\r^2\left[b_6p(\ve\cdot s)
  -\frac{b_6-b_4}{6}(R-4s^\g{_\g})\right]
  +2\ve_{ca}\r^2b_5(\ve\cdot s)\, ,                        \nn
\ee
and $\ve\cdot s:=\ve^{fg}s_{fg}$. The first piece of $\d I_{R^2}$ has
the form
\bsubeq
\be
A:=2\b_2\int_{\pd\cM}d^2x\ve^{\a\b}\left(\frac{p}{2}\ve_{ac}
   -\eta_{ac}\right)V^c\,\bare^a{_\a}\d\om_\b\, .
\ee
The second piece can be conveniently written as the sum of two terms,
$B+C$, where:
\bea
B&:=&\frac{b_6qp}{\r^2}\int_{\pd\cM}d^2x
    \ve^{\a\b}e_{a\a}\d e^a{_\b}
    -4b_6q\int_{\pd\cM}d^2x\ve^{\a\b}
            \d e_{a\a}\ve^{af}s_{\b f}                     \nn\\
  && +\d\int_{\pd M}d^2x \D_3\, ,                          \\
\D_3&:=&{4b_6q}\ve^{\a\b}\ve^{ab}e_{a\a}s_{\b b}
      -\frac{b_6q}{\r^2}\ve^{\a\b}\ve^{ab}e_{a\a}e_{b\b}
     = \frac{qb_6}{\r^2}\bare K\, .                        \nn
\eea
and
\bea
C&:=&2\int_{\pd\cM}d^2x\ve^{\a\b}\left[
  b_6p(\ve\cdot s)-\frac{b_6-b_4}{6}(R-4s^\g{_\g})\right]
  e_{a\a}\left(\frac{p}{2}\eta^{ab}-\ve^{ab}\right)\d e_{b\b}\nn\\
 && -4b_5\int_{\pd\cM}d^2x\ve^{\a\b}(\ve\cdot s)e^c{_\b}
       \left(\frac{p}{2}\eta_{ca}-\ve_{ca}\right)\d e^a{_\a}\,.
\eea
\esubeq

Now, the first terms in $\d I_{\rm EC}$, $\d I_{\rm T^2}$ and $A$ are
divergent, but their sum vanishes as a consequence of \eq{5.4a}. The
sum $I_\ct:=\int d^2x(\D_1+\D_2+\D_3)$, which appears in $\d I$
as a total variation and is also divergent, is recognized as the
\emph{counterterm}; when subtracted from $I$, it defines the
renormalized action $I_\ren=I-I_\ct$, see the next subsection for more
details. The variation of $I_\ren$ is finite:
\bea
\d I_\ren
  &=& -4a\int_{\pd\cM}d^2x\ve^{\a\b}\ve^{ac}s_{\b c}\d e_{a\a}\,,\nn\\
  &&+4a_3\int_{\pd\cM}d^2x(\ve\cdot s)\ve^{\a\b}e^a{_\a}\d e_{a\b}\,,\nn\\
  &&+A-4b_6q\int d^2x\ve^{\a\b}\ve^{af}s_{\b f}\d e_{a\a}+C\,.
\eea
From this result, one can identify the spin and the energy-momentum
boundary currents, or equivalently, the 1-point functions of an
effective 2D quantum theory, as:
\bsubeq\lab{5.11}
\bea
\s^\b&=&(b_4-b_5)\ve^{\b\a}\left(\frac{p}{2}\ve_{ac}
            -\eta_{ac}\right)V^c\,\bare^a{_\a}\, ,         \\
\t^\a{_a}&=&4(a+b_6q)\ve^{\a\b}\ve_{ac}s_\b{^c}
   +4a_3(\ve\cdot s)\ve^{\a\b}e_{a\b} \nn\\
&&-\ve^{\a\b}\frac{b_6-b_4}{3}(R-4s^\g{_\g})
  e^b{_\b}\left(\frac{p}{2}\eta_{ba}-\ve_{ba}\right) \nn\\
&&+2\ve^{\a\b}\left[ \left(b_6\frac{p^2}{2}-2b_5\right)\eta_{ba}
   +p(b_5-b_6)\ve_{ba}\right] e^b{_\b}(\ve\cdot s)\, .
\eea
\esubeq

\subsection{Renormalized action}

Before we continue to examine the Noether--Ward identities of the
boundary currents \eq{5.11},  let us stress that the variation of the
full action $I$ contains the total variation of the divergent term
$I_{\rm ct}$, which can be compactly expressed as
\bsubeq
\be
I_{ct}=\frac{(a+qb_6)}{\r^2}\int_{\pd M}d^2x\bare K
      =(a+qb_6)\int_{\pd M}d^2x\te K\, .  \lab{5.13a}
\ee
Note that the factor $(a+qb_6)$ is proportional to the central
charge of the theory \cite{x12}. Subtracting this counterterm
from the original action $I$ yields the renormalized action,
\be
I_\ren=I-I_{ct}=I-(a+qb_6)\int_{\pd M}d^2x\te K\,,
\ee
\esubeq
the variation of which produces the finite boundary currents \eq{5.11}.

One should observe that here, like in the MB model or GR, the
counterterm is of the Gibbons--Hawking type, but with a modified factor
which involves the $\hR^2$ coupling constant $b_6$. All the other
quadratic terms in the action give finite contributions that need not
be regularized. Similarly as in the previous section, we can decompose
$I_\ct$ into the Balasubramanian--Kraus type local counterterm and the
finite term proportional to $\int d^2x\bare s^c{_c}$, which becomes, on
shell, a local function of the boundary curvature and torsion.

We would like to emphasize that, in even boundary dimensions, there is a
logarithmic term in the field expansions related to the variation of the
conformal anomaly, i.e., to its functional derivative with respect to the corresponding source.
In Einstein-Hilbert gravity, however, its coefficient is
obtained as a variation with respect to the boundary metric of the conformal anomaly which is topological invariant in
two dimensions, such that it can be dropped out in the holographic
renormalization procedure \cite{Skenderis-Rees}. For the present holographic analysis with
torsion, the field equations can be also solved consistently without
adding such type of terms. This seems to be a reflection of
the fact that the coefficients of the log terms in both the vielbein
and the spin connection are related to the variation of the Weyl anomaly that
turns out to be, as we show below, a topological invariant,
even when the torsional degrees of freedom are taken into account.

Similar type of a logarithmic term also appears in the action evaluated on-shell.
Namely, the counterterm \eq{5.13a} ensures a differentiable and finite \emph{variation}
of the action $I_\ren$, but the action itself contains a log term whose coefficient
is related to topological invariants. As mentioned in Sec.4, inclusion of these terms is important
in the full renormalized action that is identified with the free energy of the dual CFT.

These invariants are the same as those appearing in the conformal anomaly,
the form of which will be obtained in the next subsection.

\subsection{Boundary symmetries and anomalies}

To simplify the derivation of the boundary symmetries and make it more
direct, we rewrite the spin and the energy-momentum current in a more
compact way. First, using the expression \eq{5.8}, we write the spin
current in the form
\be
\s^\b=-2\ve^{\a\b}\cH_{\a 1}
     =2\ve^{\b\a}\bigl(\he^a{_\a}\cH_{a 1}\bigr)|_{\r=0}\,.\lab{5.13}
\ee
In what follows, we shall omit the sign $|_{\r=0}$ for simplicity.
After isolating the counterterm, the energy-momentum tensor becomes
\bsubeq
\bea
\t^\b{_b}&=&4(a+qb_6)\ve^{\a\b}\ve_{cb}s_\a{^c}
   -\frac{2a_3}{\r^2}\ve^{\a\b}e_{b\a}\bigl(\hcA-p\bigr)   \nn\\
  &&-\frac{2}{\r^2}\ve^{\a\b}e^c{_\a}\bigl(
       \cH_{cg}-\eta_{cg}\cH-\eta_{cg}b_6q\bigr)
       \left(\frac{p}{2}\d^g_b-\ve^g{_b}\right)\, .
\eea
Then, using \eq{5.4a}, we obtain an equivalent form of $\t^\b{_b}$:
\bea
\t^\b{_b}&=&-\frac{2}{\r^2}(a+qb_6)
                           \ve^{\a\b}(k_{b\a}-\ve_{ab}e^a{_\a})
   -\frac{2a_3}{\r^2}\ve^{\a\b}e_{b\a}\hcA                 \nn\\
  &&-\frac{2}{\r^2}\ve^{\a\b}e^c{_\a}\bigl(
       \cH_{cg}-\eta_{cg}\cH-\eta_{cg}b_6q\bigr)
       \left(\frac{p}{2}\d^g_b-\ve^g{_b}\right)\, .
\eea
\esubeq
Note that the trace of $\t^\a{_a}$ is given by
\be
\t=\bar e\left[-4(a+b_6q)s^a{_a}-\frac{2}{\r^2}
  \left(\cH^a{_a}-2\cH-2b_6q
        +\frac{p}{2}\ve^{ab}\cH_{ab}\right)\right]\, .
\ee

\prg{Lorentz invariance.} To verify the conservation law of the spin
current \eq{5.13}, we start from the relations:
\bea
\nab_\a\s^\a&=&\hnab_\a\s^\a
  =\frac{\bare}{\r^2}\ve^{bc}\left(\hT^a{}_{bc}\cH_{a1}
                    +2\hnab_b\cH_{c1}\right)\, ,           \nn\\
\ve^{ab}\t_{ab}&=&
  -\frac{2\bare}{\r^2}\bigl(a+b_6q+2a_3\bigr)(\hcA-p)
  -\frac{2\bare}{\r^2}\left(\ve^{bc}\cH_{bc}+\frac{p}{2}\cH^c{_c}
                       -p\cH-pb_6q\right)                  \nn\\
 &\equiv&\frac{\bare}{\r^2}(a+2a_3-\cH^c{_c})\ve^{ab}\hT_{1ab}
   =-\frac{2\bare}{\r^2}(a+2a_3-\cH^c{_c})\hcA\, .         \nn
\eea
Then, using the field equation $(1ab)$ in the form
\be
-2\bigl(a-\cH^c{_c}+2a_3\bigr)\hcA
  +\ve^{bc}(\hT^a{}_{bc}\cH_{a1}+2\hnab_b\cH_{c1})=0\, ,   \nn
\ee
the Lorentz invariance condition is found to be satisfied on shell:
\be
A_L\equiv\nab_\a\s^\a+\ve^{ab}\t_{ab} {=} 0\, .
\ee
Thus, our parity-invariant model \eq{5.2} is Lorentz-invariant, in
contrast to the situation in the MB model, where the Chern-Simons term
violates this invariance, see \eq{4.12}.

\prg{Translation invariance.} Let us now examine the invariance under
local translations. First, we note that the validity of the Lorentz
invariance condition \eq{3.3b} implies that the last term on the
right-hand-side of \eq{3.3a} vanishes. Next, we calculate the
divergence of the energy-momentum tensor:
\bea
\nab_\b\t^\b{_a}&=&\frac{2\bare}{\r^2}
  \left[(a+b_6q)\left(\frac{1}{\r}\hR_{1a}
        -\left(\frac{p}{2}\ve_{ab}-\eta_{ab}\right)V^b\right)
        +2\nab^b\cH_{a1b}-a_3(p-\hcA)\ve_{ab}V^b\right]    \nn\\
&&+\frac{\bare}{\r^3}\left(\ve_a{^b}+\frac{p}{2}\d^a_b\right)
   \left[2\ve^{cd}\hat\nab_c(\cH_{db}-\eta_{db}\cH)
         +2\hcA\cH_{1b}+2\hcA\cH_{b1}-2k_b{^d}\cH_{d1}\right.\nn\\
&&+\left.\hT^f{}_{cd}\ve^{cd}\left(\cH_{fb}-\eta_{fb}\cH
                            -b_6q\eta_{fb}\right)\right]\,.\nn
\eea
Making use of \eq{5.4a} and the $(abc)$ field equation (Appendix \ref{expansion}),
the above result is simplified:
\bea
\nab_\b \t^\b{_a}& {=} &\frac{2\bar e}{\r^2}
  \left[(a+b_6q)\frac{1}{\r}\hR_{1a}+2\nab^b\cH_{a1b}
        -\a_3\left(\frac{p}{2}\ve_{ab}+\eta_{ab}\right)V^b\right.\nn\\
&&\left.+a_3(\hcA+p)\ve_{ab}v^b+\frac{1}{\r}
  \left(\ve_{ab}+\frac{p}{2}\eta_{ab}\right)(\hcA\cH^b{_1}
  -k^{db}\cH_{d1})\right]\,.
\eea
Then, using the relations
\bea
\s^\b F_{\a\b}&=&\bare R\cH_{\a 1}\, ,                     \nn\\
\t^{cb}T_{bac}&=& 4\bar e(a+b_6q)s_{ab}V^b
         +\frac{2\bare}{\r^2}a_3(p-\hcA)\ve_{ab}V^b        \nn\\
&&-\frac{2\bare}{\r^3}
       (\cH_{ac}-\eta_{ac}\cH-\eta_{ac}b_6 q)\hR_1{^c}\, ,
\eea
we finally obtain:
\bea
A_T&=&\nab_\b \t^b{_a}-\s^\b F_{a\b}-\t^{bc}T_{cab} \nn\\
 &=& \frac{\bare}{\r^2}\left[-4\r^2(a+b_6q+\a_3)s_{ab}V^b
  -\r R\cH_{a1}+\frac{1}{\r}(2\hR_{11}+\hR)\cH_{a1}\right. \nn\\
 && +\left.\frac{2}{\r}\left(\ve_{ab}+\frac{p}{2}\eta_{ab}\right)
   (\hcA\cH^b{_1}-k^{db}\cH_{d1})\right]                   \nn\\
 &=&-4\bar e s_a{^b}\left[\left(a+b_6q+\a_3
   -\frac{b_4-b_5}2\left(1+\frac{p^2}4\right)\right)V_b
   +p\frac{b_4-b_5}2\ve_{bc}V^c\right] {=} 0\, ,
\eea
where, in the last line, we again used the $(abc)$ field equation.

This proves the translation invariance on the boundary.

\prg{Conformal anomaly.} Let us now examine the dilatation invariance
by calculating the expression $A_C=\t-\nab_\b(\ve_{ab}\s^a e^{b\b})$. We
start with
\bea
A_C&=&
  \bare\left[-4(a+b_6q)s^c{_c}+4p(b_5-b_6)(\ve\cdot s)
  +\frac{2}{3}(b_6-b_4)(R-4s^c{_c})\right]                 \nn\\
&&+(b_4-b_5)\nab_\b\left[\bar e\left(\frac p2\ve_{ab}
                      -\eta_{ab}\right)e^{a\b}V^b\right]\,.\nn
\eea
Then, the identity
$$
\nab_\b\left[\bare\left(\frac{p}{2}\ve_{ab}-\eta_{ab}\right)
                   e^{a\b}V^b\right]
  =\bare\left[\left(\frac{p}{2}\ve_{ab}-\eta_{ab}\right)\nab^a V^b
    +V^a V_a\right]\,,
$$
and the 2nd order piece of equation $(1ab)$, lead to:
\bea
A_C&=&\bare\Bigl[-(a+b_6q)R+4b_6pq(\ve\cdot s)
             +\Bigl(a+q\frac{b_6+2b_4}3\Bigr)(R-4s^c{_c})\nn\\
&& -(q+2)(b_4-b_5)(\nab_a V^a-V_aV^a)
        +p(b_4-b_5)\ve^{ab}\nab_a V_b)\Bigr]\, .           \nn
\eea
Finally, by using equations $(1a)$ and $(11)$, we obtain the conformal
anomaly:
\bea
A_C&=&-(a+b_6q)\bare R
    +\left[2\a_3-(q+2)(b_4-b_5)\right]\bare(\nab_aV^a-V_aV^a)\nn\\
    &&+p(b_4-b_5)\bare\ve^{ab}\nab_a V_b\, .
\eea
Since the conformal symmetry is broken, the boundary symmetry is
reduced to the local Poincar\'e invariance.

The first term in $A_C$, proportional to $\bare R=\pd_\a
(2\ve^{\a\b}\om_\b)$, is a topological density (related to the
topological invariant $\int d^2x\bare \bar R$); the related
factor $(a+b_6q)$ is proportional to the central charge of the
theory \cite{x12}. Since the Weyl weights of
$e^a{_\a},T^a{}_{\b\g}, V^a,\nab_a V^a$  are $+1,+1,-1,-2$,
respectively, the remaining two terms in $A_C$ are seen to be
invariant under local dilatations. For details of the
classification of conformal anomalies, see \cite{x28}.

A closer inspection of the Weyl invariants leads to the
identities:
\bea
&&W_1:=\bare(\nab_a V^a-V_a V^a)
      =\pd_\a(\ve^{\a\b}e_{a\b}\ve^{ab}V_b)\,,\nn\\
&&W_2:=\ve^{ab}\nabla_a V_b
      =\pd_\a(\ve^{\a\b}e^a{_\b}V_a)\, .                   \lab{5.22}
\eea
In particular, the first identity can be written in the language of
differential forms as
\be
N'\equiv T^a\,{}^*T_a-e^a\nabla\,{}^*T_a=d(e^a\,{}^*T_a)\, ,\lab{5.23}
\ee
where we used $V_a=\ve_{ab}{}^*T^b$. The 2-form $N'$, which
represents $W_1$, has an interesting resemblance with the
Nieh-Yan 4-form \cite{x29,x30}. Similarly, a Nie--Yan-like
representation for $W_2$ is obtained by the replacement
${}^*T_a\to T_a$ in \eq{5.23}. The integrals of $W_1$ and $W_2$
over the boundary are topological invariants, the nature of
which will be studied elsewhere.

A theory with parameters for which the conformal anomaly
vanishes is known as the critical gravity. For such a critical
choice of parameters, the bulk theory may acquire logarithmic
modes, which leads to a logarithmic CFT at the boundary. For
general properties of gravities at the critical point, see e.g.
\cite{x31}.

\section{Concluding remarks}

In this paper, we presented an analysis of the AdS/CFT correspondence
in the realm of 3D gravity with torsion, with an underlying RC geometry
of spacetime.

Starting with a suitable holographic ansatz and its consistency
condition, we found that the expected boundary symmetry is
described by local Poincar\'e transformations plus dilatations.
Based on an improved form of the Noether--Ward identities, we
first analyzed the holographic features of the MB model, where
we confirmed the results of Klemm and Tagliabue \cite{x8},
derived by a different technique. Then, turning our attention to
the more interesting case of 3D gravity with propagating
torsion, we obtained the holographic conformal anomaly, with
contributions stemming from both the curvature and the torsion
invariants. As a consequence, the boundary symmetry is reduced
to the local Poincar\'e invariance. The improved treatment of
the Noether--Ward identities, being independent of the value of
torsion on the boundary, significantly simplifies the
calculations.

An interesting problem for further study is to clarify how
torsion affects the structure of the dual CFT.
A simple approach would be to study the specific PGT sectors containing only one of the
six propagating torsion modes, with $J^P=0^\pm, 1,2$ \cite{x12}.

\acknowledgments

One of us (O.M.) would like to thank Max Ba\~nados, Gast\'on Giribet,
Julio Oliva, Jorge Zanelli and all the other participants of the
\emph{Workshop on String Theory, Gravity and Fields}, held in Buenos
Aires in October 2012, for their useful comments in the final stage of
the preparation of this manuscript. This work was supported by the
Serbian Science Foundation under Grant No.~171031 and the Chilean
FONDECYT Grants No.~1090357 and No.~1110102. O.M. thanks DII-PUCV
for support through the project No.~123.711/2011. The work of R.O. is
financed in part by the UNAB grant DI-117-12/R.

\appendix
\section{On the RC geometry in 2D} \label{geometry}
\setcounter{equation}{0}

In 2D, the Lorentz connection, which is Abelian, has only one
independent ``internal" component, $\om^{ab}{_\a}=-\ve^{ab}\om_\a$, and
the local Poincar\'e transformations of $e^a{_\a}$ and $\om_\a$ have
the form \eq{2.12}. The corresponding field strengths, the curvature
and the torsion, are given by
\bea
&&R^{ab}{}_{\a\b}=-\ve^{ab}F_{\a\b}\, ,\qquad
  F_{\a\b}:=\pd_\a\om_\b-\pd_\b\om_\a\, ,                   \nn\\
&&T^a{}_{\a\b}=\nab_\a e^a{_\b}-\nab_\b e^a{_\a}\, ,
  \qquad \nab_\a e^a{_\b}:=\pd_\a e^a{_\b}
                           -\ve^a{_c}\om_\a e^c{_\b}\, .
\eea
The Ricci tensor and the scalar curvature read:
\bsubeq
\be
R^a{_c}=-\ve^{ab}F_{cb}\, ,\qquad R=-\ve^{ab}F_{ab}\, .
\ee
As a consequence:
\be
R_{ab}=\frac{1}{2}\eta_{ab}R\, ,\qquad F_{ab}=\frac{1}{2}\ve_{ab}R\,,
\ee
\esubeq
and the Ricci tensor is always symmetric. The torsion tensor, with only
two independent components, is completely determined by its vector
piece $V_a=T^b{}_{ba}$ as
$$
T^a{}_{bc}=\d^a_b V_c-\d^a_c V_b\, .
$$
When the torsion vanishes, the connection becomes Riemannian:
\be
\tom_\a=\frac{1}{2}\ve^{ab}(c_{abc}-c_{cab}+c_{bca})e^c{_\a}\,,
\qquad c^a{}_{\a\b}:=\pd_\a e^a{_\b}-\pd_\b e^a{_\a}\, ,   \lab{A.3}
\ee
see also \eq{3.4}.

In the Gauss-normal radial foliation, the unit normal to the boundary
$\pd M$ has the form
$$
n_i=(n_1,n_a)=\frac{\he_i{^\r}}{\sqrt{-\hat g^{\r\r}}}=(1,0,0)\,,
$$
with $n^2=-1$. The extrinsic curvature (the second fundamental form) of
$\pd M$ is defined by $K_{ij}=\hnab_i n_j$. The only nonvanishing
components of $K_{ij}$ are
\be
K_{ab}:=\hnab_a n_b=-\ve_{bc}\hom^c{_a}=\ve_{cb}k^c{_a}
       =\frac{p}{2}\ve_{ab}+\eta_{ab}-2\r^2s_{ab}\,,
\ee
where we used $k^c{_a}:=k^c{_\a}e_a{^\a}$. In particular:
\bea
&&K_{(ab)}=\eta_{ab}-2\r^2s_{(ab)}\, ,\qquad
  K^b{_b}=2-2\r^2s^c{_c}\, ,                               \nn\\
&&\ve^{ab}K_{ab}=-p-2\r^2\ve^{ab}s_{ab}\equiv-\hcA\, .     \lab{A.5}
\eea
The last equation gives an interesting geometric interpretation of the
axial torsion $\hcA$. For $\hcA=0$, $K_{ab}$ reduces to the standard
Riemannian form.

\section{Residual symmetries to second order}  \label{residual}
\setcounter{equation}{0}

At the end of Section 2, we showed that the residual symmetry
group with the parameter $f(x)$, defined by \eq{2.9}, acts as local
dilatation on the leading order of the metric, $\bg_{\a\b}$.
From \eq{2.9}, we can also find the transformation rule for the second
order of the vielbein, $s^a{_\a}$, and extend the result of Section 2
to the second order of the metric, $g_{(2){\a\b}}$.

 Indeed, using the definitions
$\bg_{\a\b}=\eta_{ab}\bare^a{_\a}\bare^b{_\b}$ and
$g_{(2)\a\b}=s_{\a\b}+s_{\b\a}$, and restricting our attention to
dilatations ($f\ne 0$), we obtain:
\bea
\d_f\bar{g}_{\a\b} &=&2f\,\bar{g}_{\a\b}\,,\nn\\
\d_f g_{(2)\a\b} &=&2f\,g_{(2)\a\b}
  -2\bare_{a(\a}\bar\nab_{\b)}f^a +2f^\g \bar T_{(\a\b)\g}\, ,
\eea
where $f_\a:=\frac{1}{2}\pd_\a f$. In the limit when torsion vanishes,
this result reduces to the Penrose--Brown--Henneaux transformation
\cite{x12a,x12b}, which was derived in Riemannian GR and used to study
universal properties of trace anomalies.

\section{Field strengths and covariant momenta} \label{F}
\setcounter{equation}{0}

\subsection{Torsion and curvature} \label{R and T}

The results of this subsection are obtained using the expression
\eq{2.11} for $k^{ab}$.

In the local Lorentz basis, the torsion components are:
\bea
&&\hT^1{}_{1c}=0\, ,                                       \nn\\
&&\hT^1{}_{bc}=\ve_{ce}k^e{_b}-\ve_{be}k^e{_c}
              =-\ve_{bc}k^e{_e}\, ,                        \nn\\
&&\hT^a{}_{1c}=-\ve^{ae}\left[\ve_{ec}+
  \left(\frac{p}{2}\eta_{ec}+k_{ec}\right)\right]
  +\r e_c{^\g}\pd_\r e^a{_\g}\, ,                          \nn\\
&&\hT^a{}_{bc}=\r T^a{}_{bc}\, ,                           \lab{B.1}
\eea
and the components of curvature read:
\bea
&&\hR_{11c}=-\r^2e_c{^\g}\pd_\r\om_\g\, ,                  \nn\\
&&\hR_{1bc}=-\r^2 F_{bc}+\ve^{ed}k_{eb}k_{dc}\, ,          \nn\\
&&\hR_{a1c}=-\left(k_{ac}+\frac{p}{2}\ve_a{^b}k_{bc}\right)
               +\r e_c{^\g}\pd_\r k_{a\g}\, ,              \nn\\
&&\hR_{abc}=
  \r e_b{^\b}e_c{^\g}(\nab_\b k_{a\g}-\nab_\g k_{a\b})\, .
\eea
The Ricci tensor and the scalar curvature are calculated from the
relations:
\bea
&&\hR_{ik}=-\ve^{mn}{_i}\hR_{mnk}\, ,                      \nn\\
&&\hR=-\ve^{mnk}\hR_{mnk}=\hR^1{_1}+\hR^a{_a}\, .
\eea

\prg{Reduction.} Equation \eq{2.11}, in which $k_{ab}$ is expressed in
terms of $e^a{_\a}$, simplifies the expressions \eq{B.1} for the
torsion:
\bea
&&\hT_{1bc}=\ve_{bc}\hcA \, ,                              \nn\\
&&\hT_{a1c}=-\ve_{ac}\hcA\, ,                              \nn\\
&&\hT_{abc}=\r T_{abc}\, ,
\eea
where $\hcA$ is the axial torsion:
$$
\hat\cA:=\frac{1}{6}\ve^{ijk}\hT_{ijk}
        =p-\r\ve^{fg}e_f{^\b}\pd_\r e_{g\b}\,.
$$
Similarly, the curvature tensor reads:
\bea
&&\hR_{11c}=-\r^2 e_c{^\b}\pd_\r\om_\b\, ,                 \nn\\
&&\hR_{1bc}=\ve_{bc}q-\r^2F_{bc}
  -\ve_{bc}\frac{p}{2}(p-\hat\cA)+\r\ve_{bc}e^{g\b}\pd_\r e_{g\b}
  +Y_{bc}\, ,                                              \nn\\
&&\hR_{a1c}=-\ve_{ac}q
  +\left(\frac{p}{2}\ve_{ac}-\eta_{ac}\right)(p-\hcA)
  +\r^3\ve_{ab}e^{b\b}\pd_\r\left(\r^{-1}\pd_\r e_{c\b}\right)
  +X_{ac}\, ,                                              \nn\\
&&\hR_{abc}=\r\left(\frac{p}{2}T_{abc}-\ve_a{^f}T_{fbc}\right)
  +Z_{abc}\, ,                                             \lab{B.9}
\eea
where $Y_{ac},X_{ac}$ and $Z_{abc}$ are given by
\bea
&&X_{ac}:=
  \r^2\ve_a{^f}e_c{^\b}\pd_\r e_{b\g}\pd_\r(e_f{^\g}e^b{_\b})
  =-\eta_{ac}\r^3\pd_\r\bigl[\r^{-2}(p-\hcA)\bigr]
               -\r(p-\hcA)e_c{^\b}\pd_\r e_{a\b}\, ,       \nn\\
&&Y_{bc}
  =-\r^2\ve_{fg}e^{f\b}e^{g\g}(\pd_\r e_{b\b})(\pd_\r e_{c\g})\,,\nn\\
&&Z_{abc}=  -\r^2\ve_{af}e_b{^\b}e_c{^\g}\bigl[
   \nab_\b(\pd_\r e^{f\a}g_{\a\g})
  -\nab_\g(\pd_\r e^{f\a}g_{\a\b})\bigr]\, .
\eea
As a consequence, the Ricci tensor and the scalar curvature read:
\bea
&&\hR_{11}=\ve^{bc}\hR_{b1c}=2q-p(p-\hcA)-\r^3 e^{c\b}\pd_\r
           \left(\r^{-1}\pd_\r e_{c\b}\right)+(p-\hcA)^2\,,\nn\\
&&\hR_{1c}=-\ve^{ab}\hR_{abc}
  =\r\left(\frac{p}{2}\ve_{cb}V^b-V_c\right)
    -\ve^{ab}Z_{abc}\, ,                                   \nn\\
&&\hR_{a1}=\ve_a{^c}\hR_{11c}
          =-\r^2\ve_a{^c}e_c{^\b}\pd_\r\om_\b\, ,          \nn\\
&&\hR_{ab}=\ve_a{^c}(\hR_{c1b}-\hR_{1cb})
  =-2\eta_{ab}q +(p\eta_{ab}-\ve_{ab})(p-\hcA)             \nn\\
&&\hspace{1cm}+\r^3e_a{^\b}\pd_\r\left(\r^{-1}\pd_\r e_{b\b}\right)
   +\r^2R_{ab}-\r\eta_{ab}e^{g\g}\pd_\r e_{g\g}
   +\ve_a{^c}(X_{cb}-Y_{cb})\,,                            \nn\\
&&\hR=-6q+3p(p-\hcA)+2\r^3 e^{c\b}\pd_\r(\r^{-1}\pd_\r e_{c\b})\nn\\
&&\hspace{1cm} +\r^2 R-2\r e^{f\b}\pd_\r e_{f\b}
                 -\ve^{ac}(2X_{ac}+Y_{ac})\, ,             \nn
\eea
where
\bea
&&\ve_a{^c}(X_{cb}-Y_{cb})
  =-\ve_{ab}\r^3\pd_\r\bigl[\r^{-2}(p-\hcA)\bigr]
   -\r(p-\hcA)\ve_a{^c}e_b{^\b}\pd_\r e_{c\b}              \nn\\
&&\hspace{3cm}
   +\r^2\ve_a{^c}E^{\b\g}(\pd_\r e_{c\b})(\pd_\r e_{b\g})\,,\nn\\
&&-\ve^{ac}(2X_{ac}+Y_{ac})=-2(p-\hcA)^2
  -\r^2\ve^{bc}E^{\b\g}(\pd_\r e_{b\b})(\pd_\r e_{c\g})\,. \nn
\eea

\subsection{Covariant momenta} \label{momenta}

Here, we rely on the conditions \eq{2.8}, which imply
$X_{abc}=\cO_4=Y_{abc}$ and $Z_{abc}=\cO_3$. The calculations in
section 5 are greatly simplified if we first find the explicit form of
the covariant momenta. In the torsion sector, we have:
\bea
&&\cH_{11c}=-2\a_3\r V_c\, ,                               \nn\\
&&\cH_{1bc}=-\cH_{b1c}=4(\a_1-\a_2)\ve_{bc}\hcA\, ,        \nn\\
&&\cH_{abc}=2\r\left(2\a_1+\a_2+\a_3\right)T_{abc}\,,
\eea
and in the curvature sector, we find:
\bea
&&\cH_{11}=2(\b_1+\b_2+3\b_3)q
   -(\b_1+\b_2+3\b_3)p(p-\hcA)\nn\\
&&\hspace{1.2cm}
    -\b_3\r^2\left(R-4s^c{_c}\right)+\cO_4\, ,             \nn\\
&&\cH_{a1}=\b_2\r
  \left(\frac{p}{2}\ve_{ac}-\eta_{ac}\right)V^c+\cO_3\, ,  \nn\\
&&\cH_{1a}=\b_1\r
  \left(\frac{p}{2}\ve_{ac}-\eta_{ac}\right)V^c+\cO_3\,,   \nn\\
&&\cH_{ab}=-2(\b_1+\b_2+3\b_3)\eta_{ab}q
  +(\b_1+\b_2+3\b_3)\eta_{ab}p(p-\hcA)                     \nn\\
&&\hspace{1.2cm}-(\b_1-\b_2)\ve_{ab}(p-\hcA)
  +\frac{1}{2}\r^2(\b_1+\b_2+2\b_3)\eta_{ab}
                           (R-4\bare^{f\b}s_{f\b})+\cO_4\,.
\eea

\section{Radial expansion of the field equations} \label{expansion}
\setcounter{equation}{0}

In this Appendix, we display higher orders in $\r$ of the vacuum field
equations \eq{5.3}, which are needed in our study of the Noether--Ward
identities for 3D gravity with propagating torsion. To zeroth order in
$\r$, the content of these equations is displayed in \eq{5.4}. The
parameter $q$ is given in \eq{2.14b} as $q=p^2/4-1$. In our
notation, $\ve\cdot s= \ve^{ab}s_{ab}$ and $\cH=\cH^k{_k}$.

\prg{(1)} Let us start by considering the components
$(ij)=(1a),(a1),(11)$  and $(ab)$ of the first field equation
\eq{5.3a}. The object $t_{ij}$ is defined in the same equation. For
each component $(i,j)$, we display first a compact form, and then the
fully expanded field equation.

\medskip
$(1a)$:
\bea
&&\hnab^m\cH_{1am}+\frac{1}{2}\cH_1{}^{mn}\hT_{amn}
  -\cH_{1a}{^n}V_n+t_{1a}=0\, ,                            \nn\\[3pt]
&&2\r\left[(2\a_1+\a_2+\a_3)+\b_1 q\right]
   \left(\frac{p}{2}\ve_{ab}V^b-V_a\right)=\cO_3\,. \label{1a}
\eea

$(a1)$:
\bea
&&\hnab^m\cH_{a1m}+\frac{1}{2}\cH_a{}^{mn}\hT_{1mn}
  -\cH_{a1}{^n}V_n+t_{a1}=0\, ,                            \nn\\[3pt]
&&-2\r\left[a+\a_3+(b_6+\b_2)q\right]V_a                   \nn\\
&&+p\r\left[a-\a_3+8(\a_1-\a_2)+(b_6+\b_2)q\right]\ve_{ab}V^b=\cO_3\,. \label{a1}
\eea

$(11)$:
\bea
&&\hnab^m\cH_{11m}+\frac{1}{2}\cH_1{}^{mn}\hT_{1mn}
  -\cH_{11}{^n}V_n+t_{11}=0\, ,                            \nn\\[3pt]
&&-2\a_3\nab_aV^a -\left[(2\a_1+\a_2-\a_3)+\b_1 q\right]V_cV^c\nn\\
&&+\left[a-\left(2\b_3-\frac{b_6}{2}\right)q\right](R-4s^\g{_\g})\nn\\
&&-2p\left[a+4(\a_1-\a_2)-b_6q\right](\ve\cdot s)=\cO_2\, . \label{11}
\eea

$(ab)$:
\bea
&&\hnab^m\cH_{abm}+\frac{1}{2}\cH_a{}^{mn}\hT_{bmn}
  -\cH_{ab}{^n}V_n+t_{ab}=0\, ,                          \nn\\[3pt]
&&2(2\a_1+\a_2+\a_3)\nab^c T_{abc}
  -\left[(2\a_1+\a_2+\a_3)+\b_1 q\right]\eta_{ab}V_cV^c     \nn\\
&&-\eta_{ab}\left(\b_3-\frac{3b_6}{4}\right)q(R-4s^\g{_\g}) \nn\\
&&+2\eta_{ab}\left[a+4(\a_1-\a_2)-b_6 q \right]p(\ve\cdot s)\nn\\
&&-4\ve_{ab}\left[a+4(\a_1-\a_2)-(\b_1-\b_2-b_6)q \right]
            (\ve\cdot s)=\cO_2\,.
\eea

\prg{(2)} Now, we turn to the components $(kij)=(a1b),(11b),(1ab)$ and
$(cab)$ of the second field equation \eq{5.3b}.

\medskip
$(a1b)$:
\bea
&&2\hT^c{}_{1b}\left(a\eta_{ca}+\cH_{ca}-\eta_{ca}\cH\right)
  +2\hnab_1(\cH_{ba}-\eta_{ba}\cH)-2\hnab_b\cH_{1a}
  -\ve_{bc}\ve^f{_a}(\cH_{1f}{^c}-\cH_{f1}{^c})=0\, ,      \nn\\[3pt]
&&-2\b_1\left(\frac{p}{2}\ve_{af}-\eta_{af}\right)\nab_bV^f
  +2p\eta_{ab}\bigl[2b_6+(\b_1-\b_2)\bigr](\ve\cdot s)     \nn\\
&&-4\ve_{ab}\left[a+4(\a_1-\a_2)+b_6(q+p^2/2)
         +(\b_1-\b_2)\right](\ve\cdot s)                   \nn\\
&& -\left(\eta_{ab}-\frac{p}{2}\ve_{ab}\right)
  \left(\frac{3b_6}{4}-\b_3\right)(R-4s^\g{_\g})=\cO_2\, . \label{a1b}
\eea

$(11b)$:
\bea
&&2\hT^c{}_{1b}\cH_{c1}+2\hnab_1\cH_{b1}
  -2\hnab_b(\cH_{11}-\eta_{11}\cH^k{_k})
  +\ve_{bc}\ve^{fg}\cH_{fg}{^c}=0\, ,                      \nn\\[3pt]
&&\left[-\left(2\a_1+\a_2+\a_3\right)-\b_1 q\right]\r V_b=\cO_3\,. \label{11b}
\eea

$(1ab)$:
\bea
&&2\hT^1{}_{ab}\left(a\eta_{11}+\cH_{11}-\eta_{11}\cH\right)
  +2\hT^c{}_{ab}\cH_{c1}+4\hnab_{[a}\cH_{b]1}
  +\ve_{ab}\ve^{fg}\cH_{fg}{^1}=0\, ,                      \nn\\[3pt]
&&4\left[a+4(\a_1-\a_2)
  +b_6 (q+p^2/2)-2(\b_1-\b_2)\right](\ve\cdot s)           \nn\\
&&  +p\b_2 V_c V^c
  -2\b_2\left(\frac{p}{2}\eta^{fg}-\ve^{fg}\right)\nab_f V_g
  -p\left(\frac{b_6}{4}+\b_3\right)(R-4s^\g{_\g})=\cO_2\, . \label{1ab}
\eea

$(cab)$:
\bea
&&2\hT^f{}_{ab}(a\eta_{fc}+\cH_{fc}-\eta_{fc}\cH)
  +2\hT^1{}_{ab}\cH_{1c}+4\hnab_{[a}(\cH_{b]c}-\eta_{b]c}\cH)
 -\ve_{ab}\ve^f{_c}\cH_{1f}{^1}=0\, ,                      \nn\\[3pt]
&&\left[a+b_6 q-\b_2\left(1+\frac{p^2}{4}\right)
  +\a_3\right]T_{cab}-\b_2p\,\ve_{ab}V_c=\cO_2\, .  \label{cab}
\eea


\end{document}